\documentstyle[12pt]{article}
\def \app{A_{\pi \pi}}

\def \beq{\begin{equation}}
\def \btd{\bar b \to \bar d}
\def \bts{\bar b \to \bar s}

\def \eeq{\end{equation}}
\def \ok{\overline{K}}

\begin{document}
\rightline{CERN-TH-96/38}
\rightline{EFI-96-05}
\rightline{TECHNION-PH-96-03}
\rightline{hep-ph/9602335}
\rightline{February 1996}
\bigskip
\bigskip
\centerline{{\bf $V_{td}$ FROM HADRONIC TWO-BODY $B$ DECAYS}
\footnote{To be submitted to Phys.~Lett.~B.}}
\bigskip
\centerline{\it Michael Gronau}
\centerline{\it Department of Physics}
\centerline{\it Technion -- Israel Institute of Technology, Haifa 32000,
Israel}
\medskip
\centerline{and}
\medskip
\centerline{\it Jonathan L. Rosner}
\medskip
\centerline{\it Div.~TH, CERN}
\centerline{\it 1211 CH Geneva 23, Switzerland}
\smallskip
\centerline{and}
\smallskip
\centerline{\it Enrico Fermi Institute and Department of Physics}
\centerline{\it University of Chicago, Chicago, IL 60637
\footnote{Permanent address.}}
\bigskip
\centerline{\bf ABSTRACT}
\medskip
\begin{quote}
Certain hadronic two-body decays of $B$ mesons are dominated by penguin
diagrams.  The ratios of rates for several such decays, including
$\Gamma(B^0 \to \overline{K}^{*0} K^0)/\Gamma(B^0 \to \phi K^0)$,
$\Gamma(B^0 \to \overline{K}^{*0} K^{*0})/\Gamma(B^0 \to \phi K^{*0})$,
$\Gamma(B^+ \to \overline{K}^{*0} K^+)$ $/\Gamma(B^+ \to \phi K^+)$, and
$\Gamma(B^+ \to \overline{K}^{*0} K^{*+})/\Gamma(B^+ \to \phi K^{*+})$,
can provide information on the ratio of Cabibbo-Kobayashi-Maskawa (CKM)
elements $|V_{td}/V_{ts}|$ in a manner complementary to other proposed
determinations.  SU(3) breaking effects cancel in some ratios. The cases of
neutral $B$ decays are free of corrections from small annihilation terms. 
\end{quote}
\medskip
\leftline{\qquad PACS codes:  12.15.Hh, 12.15.Ji, 13.25.Hw, 14.40.Nd}
\vspace{1.5in}
\leftline{\quad CERN-TH/96-38}
\leftline{\quad February 1996}
\newpage

The charge-changing weak interactions of quarks are described in the
electroweak theory \cite{GWS} in terms of the unitary Cabibbo-Kobayashi-Maskawa
(CKM) matrix $V$ \cite{CKM}.  The elements of this matrix are fundamental
quantities in the theory. A major task of future experiments is to improve our
present knowledge of these parameters. $B$ physics experiments, from which very
useful information on two of the couplings, $V_{cb}$ and $V_{ub}$, was obtained
\cite{stone}, are expected to play in the future a crucial role towards this
goal. Measurement of $\Delta m_d$, the nonstrange neutral $B$ meson mass
difference, has provided through the box diagram mechanism the strongest
constraint at present on $V_{td}$ \cite{AL,CPreview}: 
\beq \label{eqn:range} % (1)
0.12 < |V_{td}/V_{cb}| < 0.36~.~~~
\eeq
This range is consistent with unitarity of the CKM matrix and with the observed
CP violation in the neutral $K$ meson system. Another promising way to
determine $V_{td}$, or at least to further constrain it \cite{Ali}, is through
measurement of radiative penguin decays, $B^0 \to \rho^0 \gamma$, $B^0 \to
\omega \gamma$, by comparing these rates to the measured rate of $B^0 \to
K^{*0}\gamma$ \cite{CLEOgamma}. Alternatively, this may be achieved by
improving the present lower limit on $\Delta m_s$, the strange neutral $B$
meson mass difference, or by a measurement of $K^+\to\pi^+\nu\overline{\nu}$
\cite{Kpinunu}. 

In this Letter we propose a method of measuring $V_{td}$ which is based on
comparing rates of strangeness-conserving and strangeness-changing two-body and
quasi-two-body hadronic $B$ decays to noncharmed mesons. For this purpose we
consider decays which are dominated by QCD-penguin and by electroweak penguin
operators. These operators, denoted respectively by $Q_{3-6}$ and $Q_{7-10}$
\cite{penguinops}, appear in the effective Hamiltonian for charm-conserving
hadronic $B$ decays. The CKM coefficients of these operators, dominated by the
$t$ quark contributions, are given by $V_{td}V^*_{tb}$ and $V_{ts}V^*_{tb}$,
corresponding to $\Delta S=0$ and $|\Delta S|=1$ decays, respectively. Thus, up
to SU(3) breaking corrections, the ratios of corresponding decay rates is given
approximately by $|V_{td}/V_{ts}|^2$ and could provide new measurements of
$V_{td}$. 

Previous studies of penguin-dominated hadronic $B$ decays 
\cite{early,DT,Chau,RF,Deandrea,Tanimoto} calculated rates assuming
factorization of penguin amplitudes, taking specific form factors and using
representative values for the magnitude of CKM elements. These model-dependent
calculations can serve as rough estimates. Our purpose is different. We will
treat this class of processes in a model-independent manner, and will take
ratios of $\Delta S=0$ to $|\Delta S| =1$ rates, in which a major part of the
model-dependence is expected to cancel. 

In order to perform a general analysis of two- and quasi-two-body
penguin-dominated $B$ decays, let us focus first on decays to two light
pseudoscalars, generically denoted by $B\to PP$. Assuming, at a first stage,
flavour SU(3) for the strong interactions, it is very convenient \cite{SU3} to
replace the five SU(3) invariant amplitudes describing these processes
\cite{DZ,SW,GL} (see Table 1 below) by an overcomplete set of six quark
diagrams \cite{Chau}, which we denote by $T$ (tree), $C$ (colour-suppressed),
$P$ (QCD-penguin), $E$ (exchange), $A$ (annihilation) and $PA$ (penguin
annihilation). The last three amplitudes, in which the spectator quark enters
into the decay Hamiltonian, are expected to be suppressed by $f_B/m_B$
($f_B\approx 180~{\rm MeV}$) and may be neglected to a good approximation. The
presence of higher-order electroweak penguin contributions introduces no new
SU(3) amplitudes, and in terms of quark graphs merely leads to a substitution
\cite{EWPin} 
\beq \label{eqn:combs} % (2)
T\to t\equiv T + P^C_{EW}~~,~~
C\to c\equiv C + P_{EW}~~,~~
P\to p\equiv P-{1\over 3}P^C_{EW}~~,
\eeq
where $P_{EW}$ and $P^C_{EW}$ are colour-favored and colour-suppressed
electroweak penguin amplitudes. To improve the precision of the analysis, one
can then introduce first-order SU(3) breaking corrections in the amplitudes. In
Ref.~\cite{SU3br} we showed that this may be achieved in a most general manner
through mass insertions in the above quark diagrams. 

We will use the above analysis to write the amplitudes for the few processes of
type $B\to PP$ which obtain contributions from QCD- and electroweak penguin
terms (in the combination (\ref{eqn:combs}) without any contributions from tree
($T$) and colour-suppressed ($C$) terms. For generality we include at this
point smaller annihilation-like terms and SU(3) breaking terms. $\Delta S=0$
amplitudes are denoted by unprimed quantities and $|\Delta S|=1$ processes by
primed quantities. We find (see Tables I and II in Refs.~\cite{SU3} and
\cite{SU3br}): 
$$
A(B^+\to K^+ \ok^0) = p + p_3 + A~,
$$
$$
A(B^0\to K^0 \ok^0) = p + p_3 + PA~,
$$
$$
A(B^+\to \pi^+ K^0) = p' + p'_1 + A'~,
$$
\beq \label{eqn:pdecomp} % (3)
A(B_s\to K^0 \ok^0) = p' + p'_1 + p'_2 +PA'~.
\eeq
Here $p$ and $p'$ are SU(3)-invariant amplitudes. The three SU(3)-breaking
corrections \cite{SU3br} $p'_1,~p'_2,~p_3$ are due, respectively, to a $\bts$
(rather than a $\btd$) transition, an $s$ (rather than a $u$ or a $d$)
spectator quark, and $s\overline{s}$ (rather than $u\overline{u}$ or
$d\overline{d}$) pair creation. These terms may be interpreted as form-factor
and/or decay-constant corrections if one assumes factorization for penguin
amplitudes. We neglect SU(3) breaking in the smaller $A,~PA$ ($A',~PA'$)
amplitudes. 

The penguin amplitudes $p,~PA$ and $p',~PA'$ are dominated by the $t$ quark
contribution. We will neglect small $u$ and $c$ quark terms. They can affect
the magnitude of $p/p'$ \cite{BFP} by as much as 30\% for the smallest allowed
values of $|V_{td}/V_{ts}|$, but typically by at most 10\% over most of the
allowed range. The ratios $p/p'$ and $(p+PA)/(p'+PA')$ are then given simply in
terms of the ratio of corresponding CKM matrix elements: 
\beq \label{eqn:ratio} % (4)
{p+PA\over p'+PA'} \approx {p\over p'} \approx {V_{td}V^*_{tb}
\over V_{ts}V^*_{tb}} = {V_{td}\over V_{ts}}~.
\eeq
On the other hand, the ratio $A/A'$ is given by a different CKM factor
$A/A'=V_{ud}/V_{us}$. 

Neglecting annihilation-like amplitudes and SU(3) breaking terms leads to the
approximate relations 
\beq \label{eqn:firstrel} % (5)
A(B^+\to K^+ \ok^0)\approx A(B^0\to K^0 \ok^0) ~,
\eeq
\beq \label{eqn:secrel} % (6)
A(B^+\to \pi^+ K^0) \approx A(B_s\to K^0\ok^0)~,
\eeq
\beq \label{eqn:thirdrel} % (7)
{A(B^0\to K^0 \ok^0) \over A(B_s\to K^0\ok^0)}\approx
{V_{td}\over V_{ts}}~, 
\eeq
\beq \label{eqn:lastrel} % (8)
{A(B^+\to K^+ \ok^0)\over A(B^+\to \pi^+ K^0)} \approx
{V_{td}\over V_{ts}}~.
\eeq

The amplitude equality (\ref{eqn:firstrel}) for $\Delta S=0$ transitions, which
follows from the $\Delta I=1/2$ property of the $\btd$ penguin operator, can be
used to test the magnitude of the small annihilation terms which were
neglected.  Eq.~(\ref{eqn:secrel}), for $|\Delta S| = 1$ amplitudes, is
expected to be more sensitive to the SU(3) breaking term $p'_2$. In
Eq.~(\ref{eqn:thirdrel}), relating nonstrange and strange neutral $B$ decay
amplitudes, we neglect only SU(3) breaking terms, while in
Eq.~(\ref{eqn:lastrel}) for charged $B$ mesons also $A,~A'$ must be neglected. 

To evaluate the precision of the relations
(\ref{eqn:firstrel}--\ref{eqn:lastrel}), one must estimate the relative
contributions of the neglected terms. This can be done in a model-dependent
manner, for instance by assuming factorization and specific models for form
factors \cite{DT}. A rough estimate of the $A,~A'$ terms based on their
$f_B/m_B$ suppression was obtained in Ref.~\cite{SU3br}: $A/p={\cal
O}(1/5),~A'/p'={\cal O}((1/5)^3)$. The first ratio may be an overestimate if
the annihilation amplitude $A$ is further suppressed, for instance by a
helicity argument.

SU(3) breaking terms in penguin amplitudes are generally expected to lead to no
more than 30\% corrections. In the ratios of amplitudes (\ref{eqn:thirdrel})
and (\ref{eqn:lastrel}), the numerators and denominators contain different
types of SU(3) breaking terms and it is difficult to argue for cancellation
effects in a model-independent manner. As will be shown below, such a
cancellation occurs in $B$ decays involving vector mesons in the final state. 

Let us consider quasi-two-body decays of the type $B\to PV$ and $B\to VV$,
where $V$ stands for a charmless vector meson. For completeness, and in order
to treat these processes in the above SU(3) breaking framework using quark
diagrams, we digress at this point to discuss the equivalence between a
description of these processes in terms of SU(3) reduced amplitudes and quark
graphs.

In the weak Hamiltonian governing $B$ meson decays to pairs of charmless final
states ~\cite{SU3,DZ,SW,GL}, a $\bar b$ quark undergoes a transition to one
light quark [a 3 under flavour SU(3)] and two light antiquarks ($3^*$), leading
to operators transforming as $3^*,~6$ and $15^*$. Penguin operators with the
SU(3) structure $\btd$ and $\bts$ transform only as $3^*$.  The independent
amplitudes for symmetric and antisymmetric final states \cite{DZ} composed of
two flavour octets are summarized in Table 1. 

The decays of (spinless) $B$ mesons to pairs of flavour octet pseudoscalar
mesons $P$ are characterized by the reduced amplitudes in the left-hand
(symmetric) column of Table 1, since the mesons are produced in an S-wave and
the final state is symmetric under the interchange of the two final mesons.
Decays to one pseudoscalar octet and one vector ($V$) octet involve both
columns. Decays to two vector mesons involve symmetric amplitudes for $S$- and
$D$-wave final states and antisymmetric amplitudes for $P$-wave final states. 

\renewcommand{\arraystretch}{1.3}
\begin{table}
\caption{Reduced amplitudes for hadronic decays of $B$ mesons to pairs of
charmless final states.} 
\begin{center}
\begin{tabular}{c c} \hline
Symmetric & Antisymmetric \\ \hline
$\langle 27   || 15^* || 3 \rangle$ & $\langle 10^* || 15^* || 3 \rangle$ \\
$\langle  8_S || 15^* || 3 \rangle$ & $\langle  8_A || 15^* || 3 \rangle$ \\
$\langle  8_S || ~6~  || 3 \rangle$ & $\langle 10   || ~6~  || 3 \rangle$ \\
$\langle  8_S || ~3^* || 3 \rangle$ & $\langle  8_A || ~6~  || 3 \rangle$ \\
$\langle  ~1~ || ~3^* || 3 \rangle$ & $\langle  8_A || ~3^* || 3 \rangle$ \\
\hline
\end{tabular}
\end{center}
\end{table}

The relation between graphs for $B \to PP$ decays and the reduced matrix
elements in the left-hand column of Table 1 was noted in the Appendix of
Ref.~\cite{SU3}.  A corresponding expansion is possible for $B \to PV$ and $B
\to VV$ amplitudes. The need for both columns of Table 1 in describing $B \to
PV$ decays arises from the distinction between processes in which the
spectator quark enters either $P$ or $V$.

The neglect of graphs in which the spectator quark participates in the
weak interaction was shown in Ref.~\cite{SU3} to be equivalent to two
relations between reduced matrix elements:
\beq % (9)
\langle 27   || 15^* || 3 \rangle \leftrightarrow
\langle  8_S || 15^* || 3 \rangle~~~,~~~
\langle  8_S ||  3^* || 3 \rangle \leftrightarrow
\langle   1  ||  3^* || 3 \rangle~~~.
\eeq
The reduced amplitudes which are related to one another have the same flavour
structure of the effective weak Hamiltonian.  The neglect of spectator
interactions is equivalent to forbidding contractions between the SU(3) index
of the initial spectator quark and the indices associated with the weak
Hamiltonian. One must necessarily get relations among final-state amplitudes
involving the same Hamiltonian structure. 

A corresponding set of relations can be seen for the antisymmetric amplitudes:
\beq % (10)
\langle 10^* || 15^* || 3 \rangle \leftrightarrow
\langle  8_A || 15^* || 3 \rangle~~~,~~~
\langle 10   ||  6   || 3 \rangle \leftrightarrow
\langle  8_A ||  6   || 3 \rangle~~~.
\eeq
As for the symmetric amplitudes, three independent reduced matrix elements
remain when interactions with the spectator quark are neglected.  In the
graphical approach, these correspond to tree, colour-suppressed, and penguin
amplitudes. 

When considering vector mesons we shall need to discuss the $\phi$, an
octet-singlet mixture.  Aside from the special case of $B_s \to \phi \phi$,
which we discuss separately, all additional amplitudes of interest arising from
the singlet component of the $\phi$ will involve a final-state octet. These
consist of terms $\langle  8' || 15^* || 3 \rangle$, $\langle  8' || ~6~  || 3
\rangle$, and $\langle  8' || ~3^* || 3 \rangle$, where the prime denotes an
amplitude involving one final-state singlet and one final-state octet meson.

We expect processes such as $B^+ \to \pi^+ \phi$ to be highly suppressed since
the $\phi$ must be connected to the rest of the diagram by at least three
gluons, a photon, a $Z$, or a $W^+ W^-$ pair.  The last three (colour-favored
electroweak penguin) processes \cite{RF}, to which we shall return, lead to
about 10\% corrections to the QCD-dominated penguin amplitudes to be considered
here.  If the three-gluon and colour-favored electroweak penguin processes are
assumed to be zero (a good approximation), one obtains relations between each
of the three $8'$ amplitudes noted above and those in Table 1 involving the
same Hamiltonian structure. 

In the case of $B_s \to \phi \phi$, a new amplitude of the form $\langle 1'
|| 3^* || 3 \rangle$ may be related to the others by the condition that both
$\phi$'s should be connected by quark lines either to one another (as in the
``penguin annihilation'' diagrams of Ref.~\cite{SU3}, considered here to be
small), or to the rest of the diagram. 

To form all the penguin-dominated processes of the type $B \to PV$ and $B \to
VV$, we make the following observations: 

1) In $\btd$ transitions, one must consider only those decays in
which an $s \bar s$ pair is produced from the vacuum.  The production of a $u
\bar u$ pair leads to an effective transition $\btd u \bar u$
which can also arise from tree-type processes. The production of a $d \bar d$
pair can lead to mesons containing $d \bar d$ which are impossible to
distinguish from those containing $u \bar u$ (and hence which can be produced
by colour-suppressed tree-type processes.)

2) In $\bts$ transitions, one can consider production of an $s
\bar s$ pair or a $d \bar d$ pair from the vacuum.  A $u \bar u$ pair leads
again to an effective transition which can also arise from tree-type processes.

3) In decays producing an $s \bar s$ meson, we demand that it be the $\phi$,
since the $\phi$ appears to be composed mostly of $s \bar s$ and its couplings
appear to approximately respect the Okubo-Zweig-Iizuka (OZI) rule forbidding
disconnected quark diagrams.  This rule is less likely to hold for processes
involving $\eta$ and $\eta'$ (which in any case are not pure $s \bar s$)
\cite{DGRpl,GReta}. 

Using these simple rules it is straightforward to write expressions similar to
(\ref{eqn:pdecomp}) for all $PV$ and $VV$ penguin-dominated decay modes. Here
again charged $B$ decay amplitudes involve corrections from $A$ and $A'$ terms,
while the likely smaller $PA$ and $PA'$ amplitudes contribute to nonstrange and
strange neutral $B$ decays. When neglecting these terms one obtains a set of
equalities between $B^0$ and $B^+$ amplitudes. These relations, including
Eq.~(\ref{eqn:firstrel}), are given in Table 2. The amplitude equalities, which
are free of SU(3) breaking corrections, follow from $\Delta I=1/2$ and $\Delta
I=0$ selection rules of $\btd$ and $\bts$ transitions, respectively. Note that
in all cases an $s \bar s$ pair is created in the vacuum. The equalities in
Table 2 can be used to test the small magnitude of annihilation terms which
were neglected.  In Table 2 and subsequently, we shall distinguish between
$B \to PV$ and $B \to VP$ decays by adopting the convention that the second
meson is the one containing the spectator quark.  In Refs.~\cite{DT,Chau}
the $B \to PV$ penguin amplitudes are found to be very small as a result of
model-dependent dynamical cancellations.  In these cases the colour-favored
electroweak penguin contributions involving $\phi$ production by the neutral
weak current are no longer negligible. 

One may obtain a set of relations involving $B^0$ and $B_s$ $\btd$ and $\bts$
transitions, which determine $V_{td}/V_{ts}$. These relations, given in Table 3
and containing Eq.~(\ref{eqn:thirdrel}), do not require neglect of annihilation
terms. They may be affected, however, by SU(3) breaking effects which are
implicit in the table. The $q\overline{q}$ pair created out of the vacuum in
each process is indicated in the table and represents a possible SU(3) breaking
correction term of type $p_3~(p'_3)$. (An example of appreciable form-factor
effects in comparing $d \bar d$ and $s \bar s$ production is given in
Ref.~\cite{RF}.) Other SU(3) breaking corrections are a type-$p'_1$ term which
occurs in $\bts$ transitions and a type-$p_2~(p'_2)$ which contributes to $B_s$
decays. We note that the type-$p_3~(p'_3)$ terms cancel in ratios of $\Delta
S=0$ and $|\Delta S|=1$ amplitudes, such as $A(B^0\to \ok^{*0} K^0)/A(B^0\to
\phi K^0)$ and $A(B^0\to \ok^{*0} K^{*0})/A(B^0\to \phi K^{*0})$, which are
therefore expected to provide a better measure of $V_{td}/V_{ts}$ than other
ratios. 

\begin{table}
\caption{Relations involving $B^0$ and $B^+$ decays dictated by selection rules
associated with dominance of penguin amplitudes.} 
\begin{center}
\begin{tabular}{|c|c|} \hline
$B^0$ decay       & $B^+$ decay       \\ \hline
$\ok^0 K^0$       & $\ok^0 K^+$       \\
$\ok^0 K^{*0~a}$  & $\ok^0 K^{*+~a}$  \\
$\ok^{*0} K^0$    & $\ok^{*0} K^+$    \\
$\ok^{*0} K^{*0}$ & $\ok^{*0} K^{*+}$ \\ 
$\phi  K^0$       & $\phi  K^+$       \\
$\phi  K^{*0}$    & $\phi  K^{*+}$    \\ \hline
\end{tabular}
\end{center}
\leftline{\qquad $^a$Small amplitude in some models.}
\end{table}

\begin{table}
\caption{Summary of amplitude relations between $\btd$ and $\bts$
penguin-dominated $B^0$ and $B_s$ decays.  Ratios of $\btd$ and $\bts$
processes are given by $V_{td}/V_{ts}$. The pair produced out of the vacuum
in each class of decay is indicated.  Entries indicate final states. Rows
indicate amplitudes related to one another.} 
\begin{center}
\begin{tabular}{|c|c|c|c|c|} \hline
\multicolumn{2}{|c|}{$B^0$ decays} & \multicolumn{3}{c|}{$B_s$ decays} \\
\hline
$\btd$ & $\bts$ & $\btd$ &  \multicolumn{2}{c|}{$\bts$} \\
\hline
$s \bar s$ pair   & $s \bar s$ pair & $s \bar s$ pair & $d \bar d$ pair &
$s \bar s$ pair \\
\hline
$\ok^0    K^0$    &               &                 & $K^0 \ok^0$       & \\
$\ok^0  K^{*0~a}$ &              & $\ok^0 \phi^{~a}$ & $K^0 \ok^{*0~a}$ & \\
$\ok^{*0} K^0$    & $\phi K^0$    &                 & $K^{*0} \ok^0$    & \\
$\ok^{*0} K^{*0}$ & $\phi K^{*0}$ & $\ok^{*0} \phi$ & $K^{*0} \ok^{*0}$ &
$\phi \phi$ \\ \hline
\end{tabular}
\end{center}
\leftline{\qquad $^a$Small amplitude in some models.}
\end{table}

The last line of Table 3 involves the decay $B_s \to \phi \phi$, which can
occur only in even partial waves (S and D). One must then separate even from
odd partial waves \cite{DDLR} in the other decays if one wants to compare them
with the $B_s \to \phi \phi$ rate. One can then write, for example, taking
account of identical particle effects, 
\beq % (11)
\tilde \Gamma(B_s \to \phi \phi) = 2 \tilde \Gamma(B^0 \to \phi K^{*0})|_{S
+ D}~~~,
\eeq
where $\tilde \Gamma \equiv \Gamma/$(phase space).  All other relations among
$VV$ decays hold separately for each partial wave.

For many of the $B^0$ and $B_s$ decays, the flavour of the decaying meson
cannot be ascertained from the final state because it is observed in a CP
eigenstate or can be produced from both the neutral $B$ and its
charge-conjugate (via mixing).  In all cases aside from $B^0 \to \phi K^0$ and
$B_s \to \ok^0 \phi$ useful information on $|V_{td}/V_{ts}|$ still may be
obtained from time-integrated decay rates, summed over a process and its
charge-conjugate.  One can isolate the process $B^0 \to \ok^{*0} K^0$, which is
to be compared with $B^0 \to \phi K^0$, by identifying the flavour of the
initial $B$ by tagging and that of the decaying one by the charge of the kaon
in $\ok^{*0} \to K^- \pi^+$, and taking account of the known amount of $B^0
\overline{B}^0$ mixing.  The rate for $B_s \to \ok^0 \phi$, which requires
observing time-dependent $B_s - \overline{B}_s$ oscillations, is one of those
found to be small and susceptible to colour-favored electroweak penguin terms
in some models. 

Finally, another set of relations can be obtained for $B^+$ decays, in which
the ratio of $\btd$ and $\bts$ amplitudes is given by $V_{td}/V_{ts}$.
These relations, including Eq.~(\ref{eqn:lastrel}), are summarized in Table 4
which also labels the quark pair produced from the vacuum. Again, SU(3)
breaking is more likely to affect those relations in which the quark pair
produced is not the same in the $\btd$ and $\bts$ transitions. In this respect,
the amplitude ratios $A(B^+\to \ok^{*0} K^+)/ A(B^+\to\phi K^+)$ and $A(B^+\to
\ok^{*0} K^{*+})/ A(B^+\to\phi K^{*+})$ are expected to give more precise
information on $|V_{td}/V_{ts}|$ than other ratios. These relations may be
slightly affected by contributions from annihilation amplitudes. 

\begin{table}
\caption{Summary of amplitude relations between $\btd$ and $\bts$
penguin-dominated $B^+$ decays.  Ratios of $\btd$ and $\bts$
processes are given by $V_{td}/V_{ts}$.  The pair produced out of the
vacuum in each class of decay is indicated.  Entries indicate final states.
Rows indicate amplitudes related to one another.}
\begin{center}
\begin{tabular}{|c|c|c|} \hline
$\btd$  & \multicolumn{2}{c|}{$\bts$} \\ \hline
$s \bar s$ pair      & $d \bar d$ pair & $s \bar s$ pair \\
\hline
$\ok^0 K^+$       & $K^0 \pi^+$     &                 \\
$\ok^0 K^{*+~a}$  & $K^0 \rho^{+~a}$ &                \\
$\ok^{*0} K^+$    & $K^{*0} \pi^+$  & $\phi K^+$      \\
$\ok^{*0} K^{*+}$ & $K^{*0} \rho^+$ & $\phi K^{*+}$   \\ \hline
\end{tabular}
\end{center}
\leftline{\qquad $^a$Small amplitude in some models.}
\end{table}

In order to get a feeling for the level at which (type-$p_1$) SU(3) corrections
may affect these relations, let us refer to a specific model \cite{DT} for a
calculation of the above two ratios of $B^+$ amplitudes. Assuming factorization
of penguin amplitudes, the authors find the following dependence on the $\phi$
coupling and mass:
\beq
A(B^+\to\phi K^+) \simeq {\rm Const.}~V_{ts}(g_{\phi}/m_{\phi})~,~~~
A(B^+\to\phi K^{*+}) \simeq {\rm Const.}'~V_{ts}g_{\phi}~,
\eeq
where the coupling $g_{\phi}$ is obtained from the $\phi\to e^+e^-$ rate.
Correspondingly, we then obtain in this model (cancelling form factors
which are equal to a couple of percent)
\beq
{A(B^+\to \ok^{*0} K^+)\over A(B^+\to\phi K^+)} \simeq {V_{td}\over V_{ts}}
{g_{K^*}/m_{K^*} \over g_{\phi}/m_{\phi}}~~,~~
{A(B^+\to \ok^{*0} K^{*+})\over A(B^+\to\phi K^{*+})} \simeq {V_{td}\over
V_{ts}}{g_{K^*} \over g_{\phi}}~,
\eeq
where the $K^*$ weak decay constant $g_{K^*}$ is obtained from the rate of
$\tau\to K^*\nu$. Using the measured $\phi\to e^+e^-$ and $\tau\to K^*\nu$
rates, we find 
\beq \label{eqn:specific}
{A(B^+\to \ok^{*0} K^+)\over A(B^+\to\phi K^+)} = 0.97{V_{td}\over V_{ts}}~,~~~
{A(B^+\to \ok^{*0} K^{*+})\over A(B^+\to\phi
K^{*+})} = 0.85{V_{td}\over V_{ts}}~.
\eeq
The inclusion of electroweak penguin effects \cite{RF} in which the $\phi$
couples directly to the neutral current raises the two coefficients in
(\ref{eqn:specific}) by about 10\%. That is, in this particular
model-calculation, these two ratios of $\btd$ and $\bts$ amplitudes measure
$|V_{td}/V_{ts}|$ to within better than 10\%. 

The branching ratios for the $\bts$ modes are expected to be typically of order
$10^{-5}$; $B(B^+ \to \phi K^{*+})$ may be a few times $10^{-5}$
\cite{DT,Chau}. [An exception occurs for the $B \to PV$ processes (the second
line in Tables 2--4), whose rates are unpredictably small in some models.]
Corresponding $\btd$ branching ratios are expected to be about a factor of 8 to
70 smaller, depending on where in the range (\ref{eqn:range}) $|V_{td}/V_{ts}|$
lies. A few times $10^7$ $B$'s are thus expected to provide useful information
on this ratio of CKM elements. 

In summary, we have presented a general analysis of penguin-dominated two- and
quasi-two-body hadronic $B$ decays to noncharmed pseudoscalar and vector
mesons. The ratios of rates for corresponding strangeness-conserving and
strangeness-changing processes measure the CKM ratio of elements
$|V_{td}/V_{ts}|$. We have shown that corrections from annihilation graphs are
absent in neutral $B$ decays, and certain SU(3)-breaking effects can be avoided
in some cases. This method of determining $|V_{td}|$ is complementary to other
methods proposed in the past, and contains numerous possibilities for
estimating corrections in view of the large number of relations which may be
studied. 

We thank Karen Lingel for discussions and the CERN Theory Group for a congenial
atmosphere in which part of this collaboration was carried out. This work was
supported in part by the United States -- Israel Binational Science Foundation
under Research Grant Agreement 94-00253/1, by the Fund for Promotion of
Research at the Technion, and by the United States Department of Energy under
Contract No. DE FG02 90ER40560. 
\bigskip
% \newpage

% Journal and other miscellaneous abbreviations for references
% Phys. Lett. B style
\def \ajp#1#2#3{Am.~J.~Phys.~{\bf#1} (#3) #2}
\def \apny#1#2#3{Ann.~Phys.~(N.Y.) {\bf#1} (#3) #2}
\def \app#1#2#3{Acta Phys.~Polonica {\bf#1} (#3) #2 }
\def \arnps#1#2#3{Ann.~Rev.~Nucl.~Part.~Sci.~{\bf#1} (#3) #2}
\def \cmp#1#2#3{Commun.~Math.~Phys.~{\bf#1} (#3) #2}
\def \cmts#1#2#3{Comments on Nucl.~Part.~Phys.~{\bf#1} (#3) #2}
\def \cn{Collaboration}
\def \corn93{{\it Lepton and Photon Interactions:  XVI International Symposium,
Ithaca, NY August 1993}, AIP Conference Proceedings No.~302, ed.~by P. Drell
and D. Rubin (AIP, New York, 1994)}
\def \cp89{{\it CP Violation,} edited by C. Jarlskog (World Scientific,
Singapore, 1989)}
\def \dpff{{\it The Fermilab Meeting -- DPF 92} (7th Meeting of the American
Physical Society Division of Particles and Fields), 10--14 November 1992,
ed. by C. H. Albright \ite~(World Scientific, Singapore, 1993)}
\def \dpf94{DPF 94 Meeting, Albuquerque, NM, Aug.~2--6, 1994}
\def \efi{Enrico Fermi Institute Report No. EFI}
\def \el#1#2#3{Europhys.~Lett.~{\bf#1} (#3) #2}
\def \f79{{\it Proceedings of the 1979 International Symposium on Lepton and
Photon Interactions at High Energies,} Fermilab, August 23-29, 1979, ed.~by
T. B. W. Kirk and H. D. I. Abarbanel (Fermi National Accelerator Laboratory,
Batavia, IL, 1979}
\def \hb87{{\it Proceeding of the 1987 International Symposium on Lepton and
Photon Interactions at High Energies,} Hamburg, 1987, ed.~by W. Bartel
and R. R\"uckl (Nucl. Phys. B, Proc. Suppl., vol. 3) (North-Holland,
Amsterdam, 1988)}
\def \ib{{\it ibid.}~}
\def \ibj#1#2#3{~{\bf#1} (#3) #2}
\def \ichep72{{\it Proceedings of the XVI International Conference on High
Energy Physics}, Chicago and Batavia, Illinois, Sept. 6--13, 1972,
edited by J. D. Jackson, A. Roberts, and R. Donaldson (Fermilab, Batavia,
IL, 1972)}
\def \ijmpa#1#2#3{Int.~J.~Mod.~Phys.~A {\bf#1} (#3) #2}
\def \ite{{\it et al.}}
\def \jmp#1#2#3{J.~Math.~Phys.~{\bf#1} (#3) #2}
\def \jpg#1#2#3{J.~Phys.~G {\bf#1} (#3) #2}
\def \lkl87{{\it Selected Topics in Electroweak Interactions} (Proceedings of
the Second Lake Louise Institute on New Frontiers in Particle Physics, 15--21
February, 1987), edited by J. M. Cameron \ite~(World Scientific, Singapore,
1987)}
\def \ky85{{\it Proceedings of the International Symposium on Lepton and
Photon Interactions at High Energy,} Kyoto, Aug.~19-24, 1985, edited by M.
Konuma and K. Takahashi (Kyoto Univ., Kyoto, 1985)}
\def \mpla#1#2#3{Mod.~Phys.~Lett.~A {\bf#1} (#3) #2}
\def \nc#1#2#3{Nuovo Cim.~{\bf#1} (#3) #2}
\def \np#1#2#3{Nucl.~Phys.~{\bf#1} (#3) #2}
\def \pisma#1#2#3#4{Pis'ma Zh.~Eksp.~Teor.~Fiz.~{\bf#1} (#3) #2[JETP Lett.
{\bf#1} (#3) #4]}
\def \pl#1#2#3{Phys.~Lett.~{\bf#1} (#3) #2}
\def \plb#1#2#3{Phys.~Lett.~B {\bf#1} (#3) #2}
\def \pr#1#2#3{Phys.~Rev.~{\bf#1} (#3) #2}
\def \pra#1#2#3{Phys.~Rev.~A {\bf#1} (#3) #2}
\def \prd#1#2#3{Phys.~Rev.~D {\bf#1} (#3) #2}
\def \prl#1#2#3{Phys.~Rev.~Lett.~{\bf#1} (#3) #2}
\def \prp#1#2#3{Phys.~Rep.~{\bf#1} (#3) #2}
\def \ptp#1#2#3{Prog.~Theor.~Phys.~{\bf#1} (#3) #2}
\def \rmp#1#2#3{Rev.~Mod.~Phys.~{\bf#1} (#3) #2}
\def \rp#1{~~~~~\ldots\ldots{\rm rp~}{#1}~~~~~}
\def \si90{25th International Conference on High Energy Physics, Singapore,
Aug. 2-8, 1990}
\def \slc87{{\it Proceedings of the Salt Lake City Meeting} (Division of
Particles and Fields, American Physical Society, Salt Lake City, Utah, 1987),
ed.~by C. DeTar and J. S. Ball (World Scientific, Singapore, 1987)}
\def \slac89{{\it Proceedings of the XIVth International Symposium on
Lepton and Photon Interactions,} Stanford, California, 1989, edited by M.
Riordan (World Scientific, Singapore, 1990)}
\def \smass82{{\it Proceedings of the 1982 DPF Summer Study on Elementary
Particle Physics and Future Facilities}, Snowmass, Colorado, edited by R.
Donaldson, R. Gustafson, and F. Paige (World Scientific, Singapore, 1982)}
\def \smass90{{\it Research Directions for the Decade} (Proceedings of the
1990 Summer Study on High Energy Physics, June 25 -- July 13, Snowmass,
Colorado), edited by E. L. Berger (World Scientific, Singapore, 1992)}
\def \stone{{\it B Decays}, edited by S. Stone (World Scientific, Singapore,
1994)}
\def \tasi90{{\it Testing the Standard Model} (Proceedings of the 1990
Theoretical Advanced Study Institute in Elementary Particle Physics, Boulder,
Colorado, 3--27 June, 1990), edited by M. Cveti\v{c} and P. Langacker
(World Scientific, Singapore, 1991)}
\def \yaf#1#2#3#4{Yad.~Fiz.~{\bf#1} (#3) #2 [Sov.~J.~Nucl.~Phys.~{\bf #1} (#3)
#4]}
\def \zhetf#1#2#3#4#5#6{Zh.~Eksp.~Teor.~Fiz.~{\bf #1} (#3) #2 [Sov.~Phys. -
JETP {\bf #4} (#6) #5]}
\def \zpc#1#2#3{Zeit.~Phys.~C {\bf#1} (#3) #2}

\end{document}